\documentclass[10pt,journal]{IEEEtran} 

\usepackage{hyperref}
\usepackage{booktabs}
\usepackage[linesnumbered,ruled,vlined]{algorithm2e}
\usepackage{mathtools, cases}
\usepackage{textcomp}
\usepackage{makecell}
\usepackage{amsthm, enumitem}
\usepackage{amsmath,amsfonts,amssymb}
\usepackage{cite, url}
\usepackage{verbatim}
\usepackage{bm}
\usepackage{graphicx}
\usepackage{xcolor}
\usepackage{subcaption}
\usepackage{epstopdf}
\usepackage{mathrsfs}
\usepackage{txfonts}
\usepackage{lineno}
\usepackage{multirow}
\usepackage{booktabs}
\usepackage{color}
\usepackage{multicol}
\usepackage{stfloats}
\usepackage{diagbox}
\usepackage{esint}
\usepackage{float}
\usepackage{algpseudocode}
\usepackage{empheq}
\usepackage{tabularx}

\usepackage{flushend}

\allowdisplaybreaks[4]

\title{ Token Communications: A Large Model-Driven Framework for Cross-modal Context-aware Semantic Communications}

\author{Li Qiao$^{*}$, Mahdi Boloursaz Mashhadi$^{*}$, \textit{Senior Member, IEEE}, Zhen Gao, \textit{Member, IEEE},\\ Rahim Tafazolli, \textit{Fellow, IEEE}, Mehdi Bennis,~\IEEEmembership{Fellow,~IEEE}, and Dusit Niyato,~\textit{Fellow, IEEE}
\thanks{
$^{*}$ Equal contribution. {\it (Corresponding authors: Zhen Gao, Rahim Tafazolli.)}}
\thanks{
Li Qiao and Zhen Gao are with the School of Information and Electronics, Beijing Institute of Technology, Beijing 100081, China (e-mails: \{qiaoli,gaozhen16\}@bit.edu.cn).
}
\thanks{
Li Qiao, Mahdi Boloursaz Mashhadi and Rahim Tafazolli are with 5GIC \& 6GIC, Institute for Communication Systems (ICS), University of Surrey, Guildford, United Kingdom (email: \{l.qiao, m.boloursazmashhadi, r.tafazolli\}@surrey.ac.uk). 
}
\thanks{
Mehdi Bennis is with the Centre for Wireless Communications, University of Oulu, 90014 Oulu, Finland (e-mail: mehdi.bennis@oulu.fi).
}
\thanks{
Dusit Niyato is with the School of Computer Science and Engineering, Nanyang Technological University,
Singapore 639798 (e-mail: dniyato@ntu.edu.sg).
}
}

\vspace{-0.75cm}

\begin{document}

\maketitle

\vspace{-8mm}
\begin{abstract}  
In this paper, we introduce \textit{token communications (TokCom)}, a large model-driven framework to leverage cross-modal context information in generative semantic communications (GenSC). {\color{black}TokCom} is a new paradigm, motivated by the recent success of generative foundation models and multimodal large language models (GFM/MLLMs), where the communication units are \textit{tokens}, enabling efficient \textit{transformer-based token processing} at the transmitter and receiver. In this paper, we introduce the potential opportunities and challenges of leveraging context in GenSC, explore how to integrate GFM/MLLMs-based token processing into semantic communication systems to leverage cross-modal context effectively at affordable complexity, present the key principles for efficient TokCom at various layers in future wireless networks. {In a typical image semantic communication setup, we demonstrate a significant improvement of the bandwidth efficiency, achieved by TokCom by leveraging the context information among tokens.} Finally, the potential research directions are identified to facilitate adoption of TokCom in future wireless networks. The source code is publicly available at \url{https://github.com/liqiao19/TokenCom_Code}.
\end{abstract}
\begin{IEEEkeywords}
Token communications, foundation models, multimodal large language models, generative semantic communications, transformers.
\end{IEEEkeywords}

\section{Introduction}\label{sec:intro}

Motivated by emerging applications, recent research has focused on development of efficient \textit{semantic communication (SemCom)} systems \cite{SemCom1, SemCom2}, which are mostly empowered by artificial intelligence and machine learning (AI/ML)-assisted signal processing. More recently, generative AI (GenAI) models, e.g., diffusion models, GANs, and VAEs, have proven to significantly enhance communication at the semantic-level through \textit{generative SemCom (GenSC)} \cite{Liang2024generative, Li2024generative}. The recent success of powerful generative foundation models (GFMs) and multimodal large language models (MLLMs), e.g., PaLM-E, Sora, and GPT-4o, provides ample opportunities to develop ultra-low bitrate semantic communication systems. The pre-trained nature of such models and their applicability to a vast range of synthesis tasks, can revolutionize GenSC enabling intent/task-adaptive SemCom systems empowered by pre-trained GFM/MLLMs.

\begin{figure*}[tb]
\vspace{-4mm}
\centerline{\includegraphics[scale=0.79]{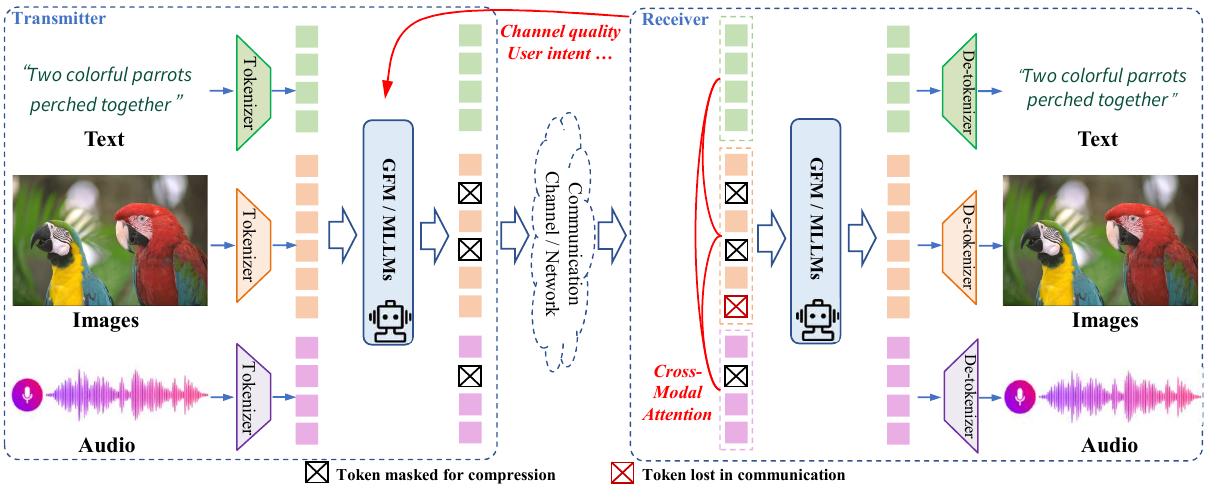}}
\captionsetup{font={footnotesize, color = {black}}, singlelinecheck = off, justification = raggedright,name={Fig.},labelsep=period}
\caption{The proposed token communications framework: Leveraging cross-modal context for efficient generative SemCom.}
\label{fig_Tokenization}
\vspace{-4mm}
\end{figure*}

Despite the above advancements, integrating cross-modal context information into semantic communication systems remains less studied. In this paper, we introduce \textit{token communications (TokCom)}, a new GenSC framework that leverages tokens as units of semantic content transmitted within the future wireless networks. Tokens are compressed representations that capture meaning through features of information-rich multimodal data, enabling efficient semantic communications. TokCom is motivated by the recent advent of powerful GFM/MLLMs based on transformer neural networks (NNs), where tokens are the basic processing units of text, images, audio/video signals that may represent words, image patches, temporal audio slices, or video subframes. {\color{black}GFMs/MLLMs}-based token processing enables to encode the semantic content of multimodal signals at the transmitter and recover the corresponding semantics at the receiver leveraging the cross-modal context information, as depicted in Fig. \ref{fig_Tokenization}. If a corrupted packet leads to an incomplete received message like ``A beach with palm [MASK] and clear blue water", where the missing word is replaced by a [MASK] token, TokCom can leverage a pre-trained LLM to predict the masked word based on the surrounding context, outputting “trees” to complete the sentence, thereby avoiding packet re-transmissions


This article aims to answer the following questions: \textit{Q1) What are the key potential opportunities and challenges of leveraging context via TokCom in GenSC? Q2) How to integrate transformer-based token processing via state-of-the-art GFM/MLLMs into SemCom systems to leverage cross-modal context effectively? Q3) What are the key principles and setups for efficient TokCom at various layers in future wireless networks?} The main contributions of this article include:
\begin{itemize}
    \item We introduce a novel TokCom framework along with its basic designs to leverage cross-modal context information in the semantic source compression, semantic channel coding, semantic multiple access, and semantic networking setups. TokCom integrates transformer-based next/masked token prediction via GFM/MLLMs. 
    \item We introduce and demonstrate a token-level loss/error mitigation scheme that leverages token likelihood estimation based on cross-modal context to predict and mitigate the tokens corrupted or lost in communication. {In a typical image semantic communication setup, we demonstrate a significant improvement of the bandwidth efficiency, by TokCom leveraging the context information.} 
\end{itemize}

\section{Tokenization, and Embedding of Various Data Modalities} \label{SEC_Proposed}
To process large multimodal data, it is first segmented into various chunks, each consisting of several \textit{tokens}. Tokens are then assigned unique IDs. Each token ID is mapped to a vector that represents the semantic meaning of the token in a dense, fixed-dimensional space, i.e., the \textit{embedding space}. The embedding vectors are learned via pre-training to capture both syntactic and semantic relationships between tokens. These embeddings allow multimodal language models process inputs more effectively by capturing token semantics as well as the relationships between different tokens, i.e., \textit{context}. A Tokenizer generates a large unified codebook of token embeddings that represent the whole corpus of multimodal data, e.g., colossal clean crawled corpus (C4) for text, LAION-5B for image-text pairs, and then the (pre-)trained MLLM essentially learns relations between tokens capturing the semantics and context. {As an example, a vision language model, e.g., the contrastive language-image pretraining (CLIP) model\footnote{\url{https://github.com/openai/CLIP}}, learns through pre-training on large corpus of image-text pairs to relate the tokens of a “book” image with its corresponding text token. Similarly, the CLAP\footnote{\url{https://github.com/LAION-AI/CLAP}} and AudioCLIP\footnote{\url{https://github.com/AndreyGuzhov/AudioCLIP}} models learn the cross-modal relations via a shared embedding space for Text-Audio and Text-Image-Audio modalities.} 



{\bf Text Tokenization/Embedding:} In text, a token can represent a word, part of a word, or even a single character, depending on the tokenization model.  Tokenization algorithms like WordPiece, byte-pair encoding (BPE), and unigram language modeling split words into subwords, with each subword mapped to a unique token in the vocabulary. Tokens are then projected into a high-dimensional feature space using word embeddings. Early models like Word2Vec and GloVe used fixed embeddings, while modern transformer models such as bidirectional encoder representations from transformers (BERT) and GPT learn contextual embeddings, meaning the representation of a word like ``bank" varies depending on its context. 

{\bf Image/Video Tokenization/Embedding:} {\color{black}Image tokens are created by dividing the image into fixed-size patches, which are then flattened and embedded. The challenge lies in balancing image fidelity and patch size; smaller patches can increase computational load, while larger ones may lose detail. For instance, vision transformer (ViT) uses $16 \times 16$ pixel patches. After patchification and embedding, vector quatization is typically used to achieve a discrete latent space, where each image is represented by a sequence of token indices from a codebook. 


{\bf Audio Tokenization/Embedding:} Audio signals are typically tokenized by transforming the waveform into spectrograms, typically using techniques such as log Mel filterbanks, which capture both the temporal and frequency-based features of the audio signal. The spectrograms are then divided into overlapping patches, and then flattened and projected into the embedding space to be processed by attention-based transformer models.

{ Finally, some commonly used tokenizers/models are BPE and WordPiece for text, VQ-VAE/GAN, TiTok\cite{yu2024image} for image, VideoMAE\footnote{\url{https://github.com/MCG-NJU/VideoMAE}} and VidTok\footnote{\url{https://github.com/microsoft/VidTok}} for video, and HuBERT\footnote{\url{https://github.com/bshall/hubert}} for Audio signal modalities.}

\begin{figure*}[tb]
\vspace{-4mm}
\centerline{\includegraphics[scale=0.83]{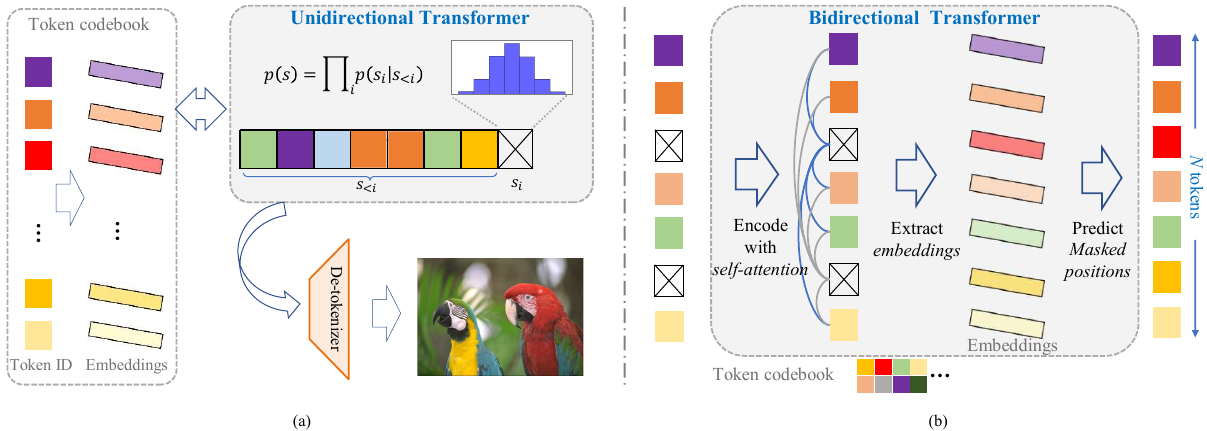}}
\captionsetup{font={footnotesize, color = {black}}, singlelinecheck = off, justification = raggedright,name={Fig.},labelsep=period}
\caption{Transformer-based context learning/prediction: (a) Unidirectional (next token prediction); (b) Bidirectional (masked token prediction).}
\label{fig_TokPred}
\vspace{-2mm}
\end{figure*}

\section{Tokens, Attention, and Transformers}\label{SEC_Proposed}

Tokens are commonly used in various vision and natural language processing (NLP) models to represent data in a structured, numerical format that models can process. Transformer-based models, e.g., ViT, BERT, and GPTs, revolutionized token processing by introducing a fundamentally different approach to handling token sequences based on the \textit{attention mechanism}. Transformers use self-attention mechanisms, where each token directly attends to all other tokens in the sequence, making it much easier for the model to capture long-range dependencies and contextual information. Compared to other NLP models, Transformers scale much more effectively \cite{kaplan2020scaling}, making it possible to train very deep models, such as the GPTs, {\color{black}LLaMA, or PaLM} with billions of parameters. Transformer-based models may learn and predict contextual information either in a unidirectional or bidirectional fashion.

\textit{Unidirectional context learning/prediction} refers to models that process input data in a single direction, e.g., left-to-right for text or row-by-row for images{\color{black}, where each token is predicted based only on the preceding tokens. Auto-regressive models like GPTs use this approach and are suited for tasks such as text completion or machine translation, where tokens are generated sequentially without access to future tokens, as depicted in Fig. \ref{fig_TokPred}(a).} 

\textit{Bidirectional context learning/prediction} leverages the concept of \textit{masking}, {\color{black}where random tokens in the input text or image are hidden, and the model learns to predict them by attending to both preceding and succeeding tokens. This bidirectional attention helps capture richer context. Models like BERT and MaskGIT \cite{MaskGIT} use masked modeling predict masked tokens by considering both nearby and distant tokens, as depicted in Fig. \ref{fig_TokPred}(b). Masked modeling is widely applied to generation tasks, such as image generation \cite{MaskGIT}, where the model iteratively generates image tokens, revealing a portion of tokens at each step and progressively uncovering more with each iteration.}

\section{Token Communications: Opportunities and Challenges} \label{OppChall}

TokCom provides the following main opportunities/advantages:
\begin{itemize}
    
    \item TokCom provides opportunities for more efficient semantic coding and communication in future wireless networks, as provided in the next sections. For example, unidirectional token prediction can be leveraged for adaptation of the coding/modulation order or the transmission power based on the next token likelihoods predicted at the transmitter by the GFMs/MLLMs. We will {\color{black}discuss} this and other basic TokCom setups in the next sections.

    \item In TokCom, the pre-trained token codebook is used as the shared knowledge base (KB) between the transmitter and receiver. This alleviates excessive knowledge sharing overheads in the conventional SemCom schemes. 
          
    \item Tokens are discrete representations that capture semantics and context, and hence, TokCom is inherently digital and more compatible with the existing multi-layer design of digital communication networks. Different from many conventional deep joint source channel coding (DeepJSCC)-based SemCom schemes, TokCom achieves scalability and adaptability by alleviating the need for end-to-end training.

    \item Tokens unify modalities and TokCom can leverages cross-modal relations to achieve ultra-low-bitrate SemCom. Furthermore, the in-context learning capabilities of MLLMs enables developing efficient TokCom frameworks that can adapt to various tasks, such as reconstruction, generation, and segmentation, {\color{black}among} different modalities.
\end{itemize}

Table \ref{CompTab} provides a comparison between TokCom and the conventional SemCom schemes. Despite the above benefits, the large computational complexity of many existing GFMs/MLLMs is a challenge to be addressed in design of efficient TokCom systems. An effective solution to this would be collaborative cloud-edge-device TokCom based on task offloading, which we further investigate in the next sections. 

\begin{table*}[htbp]
    \centering
    \renewcommand{\arraystretch}{1.3} 
    \caption{Comparison between the conventional SemCom and the proposed TokCom scheme.}
    \label{CompTab}
    \begin{tabular}{|p{3.8cm}|p{6.2cm}|p{6cm}|}
        \hline
        \textbf{} & \textbf{Conventional SemCom} & \textbf{TokCom} \\
        \hline
        \textbf{Tokenization/Embedding} & N/A & Required \\
        \hline
        \textbf{Transformer-based Signal Processing} & 
        May use transformer-based signal processing but at the pixel/sample/voxel level. & 
        Uses transformers to learn/exploit context information at the token level. \\
        \hline
        \textbf{Knowledge Base (KB)} & 
        A shared KB of various formats, e.g., graph and DNN, is required between the TX/RX. & 
        The shared KB includes a multimodal token codebook. \\
        \hline
        \textbf{Model Training} & 
        End-to-end training typically required, specifically in the JSCC setup. & 
        Leverages multimodal and typically large pre-trained models. \\
        \hline
        \textbf{Digital or Analog?} & 
        May be digital or analog (e.g., in the {\color{black}DeepJSCC} setup). & 
        Is based on a digital token codebook. \\
        \hline
        \textbf{Computational Complexity} & 
        Typically uses smaller task-specific NN models. & 
        Typically based on computationally complex {\color{black}GFMs/MLLMs}, GPTs, etc. \\
        \hline
    \end{tabular}
\end{table*}

\section{Basic Token Communication Setups} \label{SEC_Proposed}
In this section, we introduce four basic TokCom setups: (1) TokCom for semantic source compression, (2) TokCom with semantic channel coding, (3) TokCom for semantic multiple access, and (4) TokCom network protocols.


\subsection{TokCom and Semantic Source Compression}
TokCom systems can \textit{leverage {\color{black}GFMs/MLLMs}-based context processing to achieve ultra-low-bitrate semantic source compression} in future wireless networks. Language modeling is inherently a form of compression because it involves accurate prediction of the next/masked tokens in a sequence, thereby reducing uncertainty about the data. State-of-the-art GFM/MLLMs predict the conditional probabilities of tokens given context. The predictability-compression link is general across various data modalities, and by minimizing the log-likelihood loss during training, MLLMs implicitly optimize for compression. Thereby, compression is achieved when models effectively capture patterns and dependencies in the input multimodal data, making GFM/MLLMs inherently capable of functioning as efficient semantic compressors. As an example, the authors in \cite{deletang2024language} demonstrated that Chinchilla 70B, i.e., a language model primarily trained on text, can efficiently compress ImageNet patches to 43.4\% of their original size and LibriSpeech audio samples to 16.4\%, outperforming modality-specific compressors like PNG (58.5\%) and FLAC (30.3\%). {\color{black}Additionally, since tokenizers transform multimodal data into discrete tokens, tokenization inherently acts as a form of lossy compression. Its performance limits are dictated by the rate-distortion-perception theory \cite{RDP1}, which defines the trade-off between rate, distortion, and the perceptual quality of the reconstructed signal. Moreover, MLLMs can selectively discard some of the tokens at the transmitter that are less relevant to downstream tasks, enabling token-based lossy compression.} We demonstrate TokCom for semantic source compression in Fig. \ref{fig:4subfigure}(a).

\subsection{TokCom with Semantic Channel Coding}
Another basic TokCom setup involves dynamic \textit{adaptation of the channel coding and modulation scheme based on the next/masked token predictability.} In the current wireless networks, the modulation and coding scheme (MCS) index is adapted solely based on the channel quality acquired by a channel quality indicator functionality. However, in TokCom, the MCS index is adapted for each token (or block of tokens) based on both the channel quality and token predictability based on the cross-modal context information. The token likelihoods are thereby estimated by carrying out token-based context processing in conjunction with the conventional log-likelihood ratio (LLR) calculations using the channel information to determine the modulation and coding scheme. Utilizing auto-regressive context processing, i.e., next token prediction, in this setup reduces the content buffering delay, while bidirectional context processing, i.e., masked token prediction, can reduce the computational complexity. Apart from the channel coding and modulation scheme, the allocated power or other communication resources can also be optimized based on tokens' predictability. 

Another possibility is to \textit{leverage the semantic similarity of tokens to optimize a mapping from the token codebook to the channel symbols.} For example, the two tokens ``King" and ``Queen" are semantically very similar. In fact, if the gender information is lost or mistaken due to channel errors, it can be simply inferred or corrected from numerous contextual clues at the receiver leveraging GFM/MLLMs. In other words, semantically similar tokens should be mapped onto channel symbols that are more probable to be confused during transmission. State-of-the-art multimodal embedding techniques enable the semantic similarity of tokens to be measured by their distance in the embedding space. We demonstrate this concept in Fig. \ref{fig:4subfigure}(b).


\begin{figure*}[tb]
\vspace{-4mm}
\centerline{\includegraphics[scale=0.64]{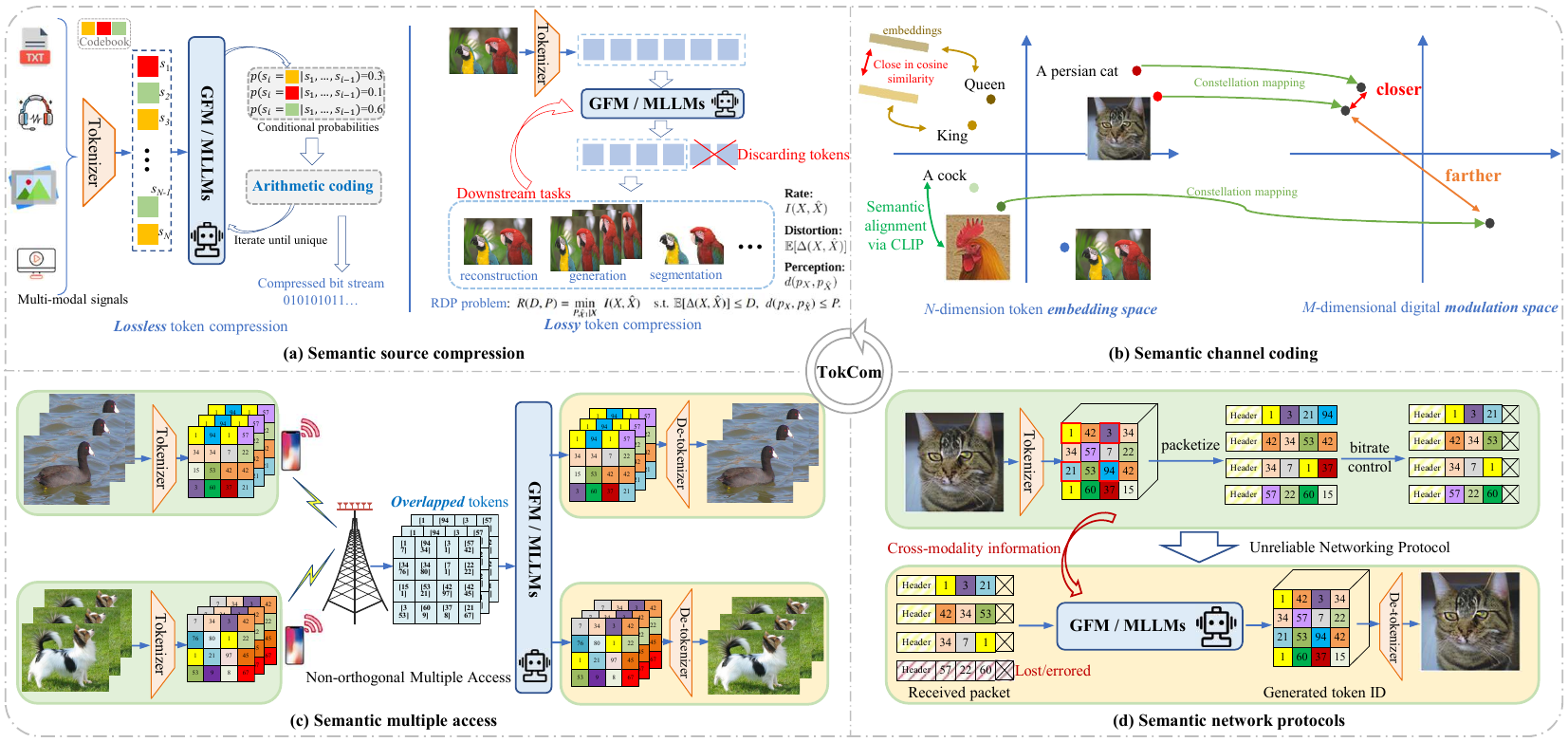}}
\captionsetup{font={footnotesize, color = {black}}, singlelinecheck = off, justification = justified,name={Fig.},labelsep=period}
\caption{Basic TokCom setups: (a) Semantic source compression: Left - lossless compression via arithmetic coding and MLLMs for probability prediction; right - discarding tokens less relevant to downstream tasks for token-based lossy compression. (b) Semantic channel coding: Map the embedding space to the modulation space; closer constellations for embedding vectors with high cosine similarity. (c) Semantic multiple access: Tokens from two users overlap in non-orthogonal multiple access, MLLMs separate sources {\color{black}using context and semantic orthogonality}. (d) Semantic network protocols: Transmitter packetizes tokens and randomly drops for bitrate control; receiver predicts lost/errored packets using MLLMs.}
\label{fig:4subfigure}
\vspace{-4mm}
\end{figure*}

\subsection{TokCom for Semantic Multiple Access}
{\color{black}GFMs/MLLMs}-based token processing can enable efficient semantic multiple access schemes in future wireless networks. To achieve this, we introduce the concept of \textit{semantic orthogonality in the token domain} as an emerging new dimension for multiple access communications. This allows several devices to transmit over the same multiple access channel (MAC). If collisions occur, i.e., signals from multiple devices are mixed non-orthogonally over the channel, a GFM/MLLM is utilized at the receiver to separate the devices’ individual signals at the token level, leveraging semantic orthogonality. In other words, GFM/MLLMs can leverage their pre-trained predictive capabilities to distinguish individual signals based on each device’s semantic context and the semantic orthogonality. 

We demonstrate this concept as an example in Fig. \ref{fig:4subfigure}(c), where two users are transmitting video streams depicting semantically orthogonal content: ``A swimming water hen." and ``A playing dog." Each video undergoes independent tokenization, with corresponding token sequences transmitted synchronously. When co-channel mixed tokens arrive at the receiver, the GFM/MLLM architecture disentangles the overlapping tokens through joint analysis of semantic orthogonality and predictive modeling, ultimately reconstructing both original videos. Building upon this principle, we recently developed token domain multiple access (ToDMA) – a practical protocol demonstrating efficient TokCom in semantic-aware multiple access scenarios \cite{qiao2025todma}.

\subsection{TokCom and Network Protocols} \label{Network}
In TokCom, the semantic information is encoded into discrete tokens that are transmitted as semantic units. Each token is transmitted by a few bits that represent its index in the token codebook, and carries efficiently compressed semantic information leveraging multimodal context. Each TokCom packet includes several tokens depending on the packet size. Although the TokCom packets are similar to the conventional packets, i.e., they are basically containers of 0s and 1s, they include strong context information for each data flow. Leveraging this context information, some level of packet loss can be efficiently mitigated at the receiver. The added robustness to packet loss allows use of less reliable network protocols (e.g., UDP) for TokCom, or further relaxation of the flow and congestion control mechanisms at various protocol layers. With TokCom, less packet re-transmissions are required, and loss of packet ordering can be mitigated at the receiver to some extent, leveraging context information. This in turn can reduce the required sequence numbers, header sizes, and consequently the overall protocol overheads. Context-aware repeat requests allows selective retransmissions only for packets with irrecoverable token errors, thereby increasing the network capacity. Context-aware routing with TokCom allows to leverage pre-trained MLLMs at the transmitter or intermediate routers and prioritize less predictable tokens. This enables adaptation of functionalities at various network layers including routing and congestion/flow control, based on multimodal token likelihoods, thereby improving the quality of service (QoS). We demonstrate the above ideas with an example in Fig. \ref{fig:4subfigure}(d).

\section{Case Study: Cross-Modality TokCom for Generative Image Semantic Communication} \label{SemMA}

We consider a semantic image transmission task leveraging the proposed TokCom framework over a wireless communication channel evaluated on the ImageNet100 dataset, where each image has a resolution of $h\times w=256\times256$ pixels.

\subsection{TokCom Setup}
Each image is first tokenized into a sequence of $N=256$ discrete tokens. To facilitate efficient transmission, these tokens are grouped into packets, with each packet containing 4 tokens. To mitigate the impact of burst errors—where consecutive erroneous tokens can significantly degrade reconstruction quality due to the loss of contextual information—the token positions included within each packet are randomized. The token codebook size is $Q=1024$ and each token is represented by 10 bits, achieving the ultra-low-bitrate transmission of 0.039 bit per pixel (bpp).


For reliable transmission, each packet undergoes rate-\(\frac{1}{2}\) convolutional encoding with cyclic redundancy check to enhance error detection and correction. The 16-QAM modulation is used and the bandwidth is 0.05 MHz. On the receiver side, demodulation and decoding are performed using a Soft Viterbi decoder, resulting in different packet error rate (PER) at different signal-to-noise ratio (SNR) values. Then, the average number of packet retransmissions required for successful data reception, denoted as $T$, is calculated as $\frac{1}{1-\text{PER}}$. {Finally, we report the token-level performance metrics, token error rate (TER) and token communication efficiency defined as $\mathrm{TCE}=\frac{h\times w}{T\times N \times \log_2({Q})}$.}

\subsection{Proposed Cross-Modality TokCom Scheme}
To enhance the TCE of TokCom under various channel conditions, we propose a ``TokCom w/ CMI" scheme, which aims to reduce the overhead of retransmissions and predict lost token packets by exploiting the context and \textit{cross-modality information} (CMI). In this scheme, each packet is transmitted only once, resulting in a retransmission count of $T=1$. This eliminates the need for packet re-transmissions, and hence, improves the TCE. For packets that experience errors, the corresponding token positions are marked with a special token, [MASK], in the decoded token sequence. The decoder then iteratively predicts these masked tokens by leveraging the contextual information surrounding them, using a pre-trained bi-directional transformer model called MaskGIT \cite{MaskGIT} with the VQGAN-based tokenizer. To further improve the prediction accuracy of the lost tokens, we incorporate CMI as a conditioning signal. Specifically, a 7-bit image class label, which is transmitted through a separate channel, is assumed to be perfectly recovered by the receiver and used as CMI. This label provides additional contextual information that helps the transformer make more accurate predictions for the lost tokens, improving quality of the received image.

{In addition to the proposed ``TokCom w/ CMI" framework, we evaluate several baseline schemes to assess the performance improvements achieved by TokCom. As TokCom enables {digital semantic communication,} we have used digital image compression/SemCom benchmarks working at similar bpp values, and all schemes use identical settings for packet length, channel coding, and modulation for a fair comparison. ``Cheng \cite{cheng2020learned}" adopts the recently developed powerful learned image compression method \cite{cheng2020learned}, with the pre-trained model fine-tuned to achieve a bpp of $\approx 0.04$. ``VQGAN \cite{esser2021taming}" employs the tokenizer but without transformer, to demonstrate the gains from transformer-based context processing in TokCom. We also simulate ``Cheng \cite{cheng2020learned} + R" and ``VQGAN \cite{esser2021taming} + R", in which the receiver requests retransmissions for any corrupted packets to guarantee accurate reception of all tokens. Finally, ``TokCom w/o CMI" represents a simplified version of TokCom that omits the CMI, serving as a baseline to quantify the performance gains due to the use of CMI.
}

\begin{figure*}[tb]
\vspace{-4mm}
\centerline{\includegraphics[scale=0.35]{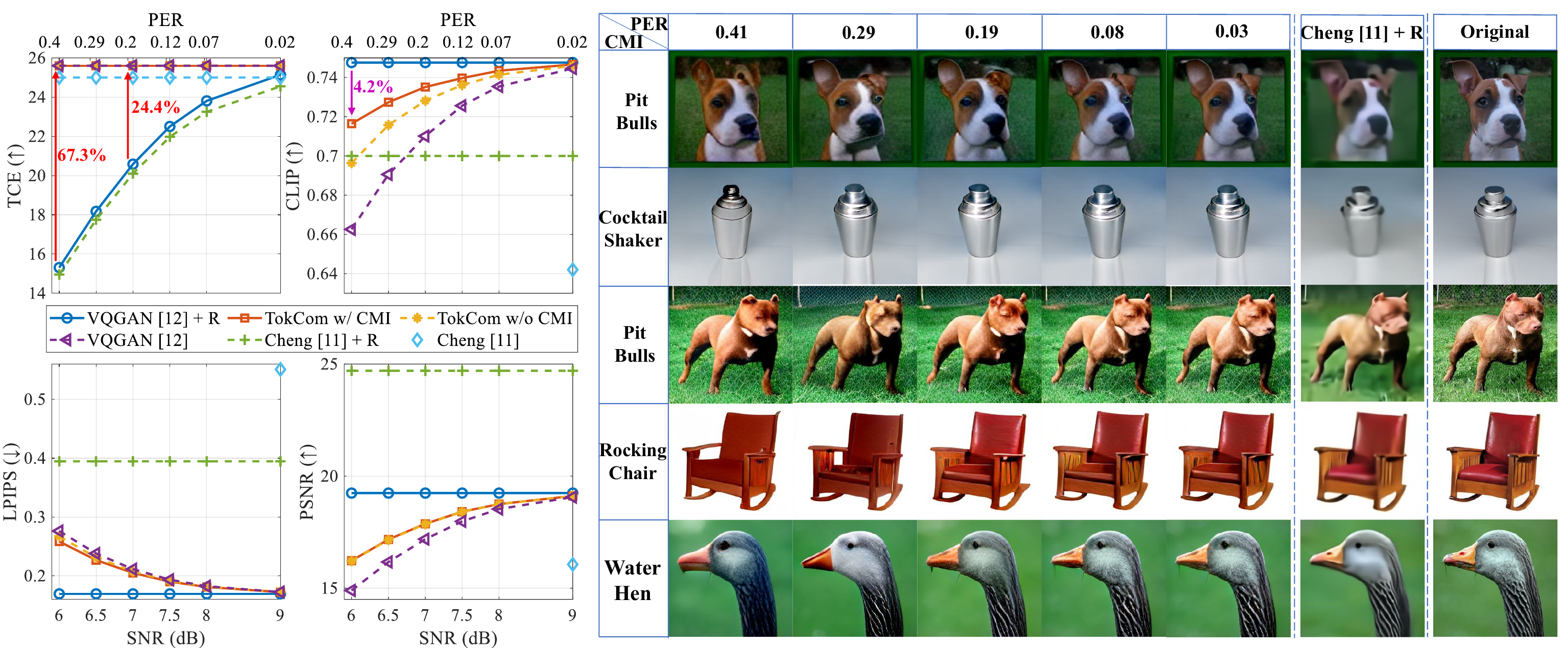}}
\captionsetup{font={footnotesize, color = {black}}, singlelinecheck = off, justification = justified,name={Fig.},labelsep=period}
\caption{Cross-modality TokCom performance for generative image semantic communication. The TCE metric represents the \textit{token communication bandwidth efficiency}. The semantic and perceptual quality are measured by CLIP (↑) and LPIPS (↓) metrics, while the pixel distortions are measured by PSNR (↑).}
\label{fig_SimResults}
\vspace{-4mm}
\end{figure*}


\subsection{Performance Analysis}
{In Fig. 4 we have compared the proposed TokCom framework with and without cross modality information (CMI) with the above benchmarks, and the results demonstrate significant gains for TokCom. Firstly, the conventional scheme, i.e., Cheng \cite{cheng2020learned} benchmark, fails to achieve acceptable reconstruction quality without retransmissions in presence of higher packet errors, i.e., the PSNR for this benchmark drops below 10dB when PER$~>~$0.07 or SNR$~<8$\,dB, which leads to unacceptable visual quality. On the other hand, the VQGAN \cite{esser2021taming} benchmark provides the performance closer to that of TokCom, but the performance gap in the quality metrics is considerable as the PER grows, specifically in terms of the CLIP metric that demonstrates the semantic quality. This highlights the gains from transformer-based context processing that enables mitigating errors without requiring retransmissions. On the other hand, when we apply retransmissions in the ``Cheng \cite{cheng2020learned} + R" and ``VQ-GAN \cite{esser2021taming} + R" benchmarks, the TCE drops severely specifically for higher PER values. Notably, the ``Cheng \cite{cheng2020learned} + R" scheme generates significantly blurrier images under the same low bpp setting and delivers inferior semantic and perceptual quality compared to TokCom, as evidenced by both CLIP and LPIPS metrics. Another interesting observation is that VQ-GAN \cite{esser2021taming} performs better than Cheng \cite{cheng2020learned} in terms of the CLIP and LPIPS metrics which better demonstrate the semantic/perceptual quality, although Cheng \cite{cheng2020learned} performs better in terms of the PSNR which is more of a distortion metric and of less importance in SemCom.

Furthermore, a comparison between the two TokCom variants reveals that the ``TokCom w/o CMI" version performs worse, with lower CLIP scores, when compared to ``TokCom w/ CMI". This validates the importance of incorporating CMI in preserving semantic quality. As shown in the visual samples of Fig.~\ref{fig_SimResults}, the class label ``Pit Bulls" serves as CMI to guide the lost image token prediction, enabling the receiver to reconstruct a Pit Bull with high perceptual quality, even with a large PER. The performance gap between these variants underscores the vital role that CMI plays in improving prediction accuracy and preserving image semantics. We specifically see that leveraging cross-modal context information to avoid retransmissions increases the bandwidth efficiency with negligible loss of semantic/perceptual quality. At a moderate PER of $20\%$, the system achieves a substantial $24.4\%$ improvement in TCE, while maintaining a CLIP score comparable with the conventional baseline. This demonstrates that TokCom not only improves bandwidth efficiency but also preserves the semantic quality of the transmitted image. When the channel conditions degrade to an SNR of $6$\,dB, resulting in a higher PER of $40\%$, TokCom exhibits remarkable robustness. The system experiences only a slight $4.2\%$ degradation in CLIP score, maintaining a score above $0.7$, which highlights its ability to handle more challenging communication environments without significant semantic loss. Additionally, as shown in Fig.~\ref{fig_SimResults}, the generative token prediction capability of TokCom ensures that both LPIPS and PSNR degrade gradually with increasing PER, similar to the trend observed in CLIP. This demonstrates that TokCom consistently preserves high perceptual quality even under challenging error conditions. Finally, while incorporating CMI leads to improvements in semantic and perceptual quality metrics, i.e., CLIP and LPIPS, its impact on distortion metrics like PSNR is minimal. This underscores the importance of semantic and perceptual quality metrics in evaluating generative semantic communication, as they more accurately reflect the quality of reconstructed content in this context. Finally, the received TER values before and after transformer-based context processing are [0.15, 0.1, 0.065, 0.038, 0.02, 0.005] and [0.03, 0.025, 0.019, 0.0126, 0.00764, 0.0021], respectively, corresponding to SNRs of [6, 6.5, 7, 7.5, 8, 9] dB. The result confirms effectiveness of transformer-based context processing in reducing the TER.}

{In terms of the communication/computation latency tradeoffs, we have used the VQGAN and MaskGIT for tokenization and context processing in simulations of Fig. 4, which requires a total of $\approx 0.8$ TFLOPs per image. Considering two recent edge AI chips, NVIDIA DGX Spark and Jetson AGX Orin with specifications 1000 FP4 TOPS and 275 INT8 TOPS (1 TFLOP (FP32) $\approx$ 12 TOPS (INT8) $\approx$ 24 TOPS (FP4)), the computation time per image is 19.2 and 34.9 msec, respectively. On the other hand, considering the bpp $\approx 0.04$ and the modulation/coding used, as well as the average number of retransmissions for each SNR/PER value, the communication time per image is [43.7, 36.9, 32.8, 29.8, 28.2, 26.7] msec for the SNRs [6, 6.5, 7, 7.5, 8, 9] dB, respectively. Considering the rapid growth in AI computation hardware, the above results provide strong evidence that additional computation time should be spent to run the large models in TokCom in order to avoid retransmissions and save on the communication time in more adverse channel/network conditions when the PER is high.}

\subsection{Other Modalities, Datasets, and Tokenizers}

{Fig. \ref{fig_SimResults5} presents examples of TokCom on Flicker image\footnote{\url{https://www.kaggle.com/datasets/hsankesara/flickr-image-dataset}} and ESC-50 audio\footnote{\url{https://github.com/karolpiczak/ESC-50}} datasets incorporating audio and text CMI using another SOTA tokenizer and transformer context processing model, i.e., TiTok\footnote{\url{https://github.com/bytedance/1d-tokenizer/blob/main/README_MaskGen.md}}. The CLAP and CLIP models are used for audio and text CMI, respectively. The token codebook size is $Q=8192$, and each image is represented by 128 tokens, resulting in a bitrate of 0.025 bpp. At the receiver side, masked token prediction modeling is used to leverage the CMI for recovering corrupted tokens. These  results show applicability of the proposed TokCom framework to other modalities, datasets, and tokenizer models. }

\begin{figure}[tb]
\centerline{\includegraphics[scale=0.425]{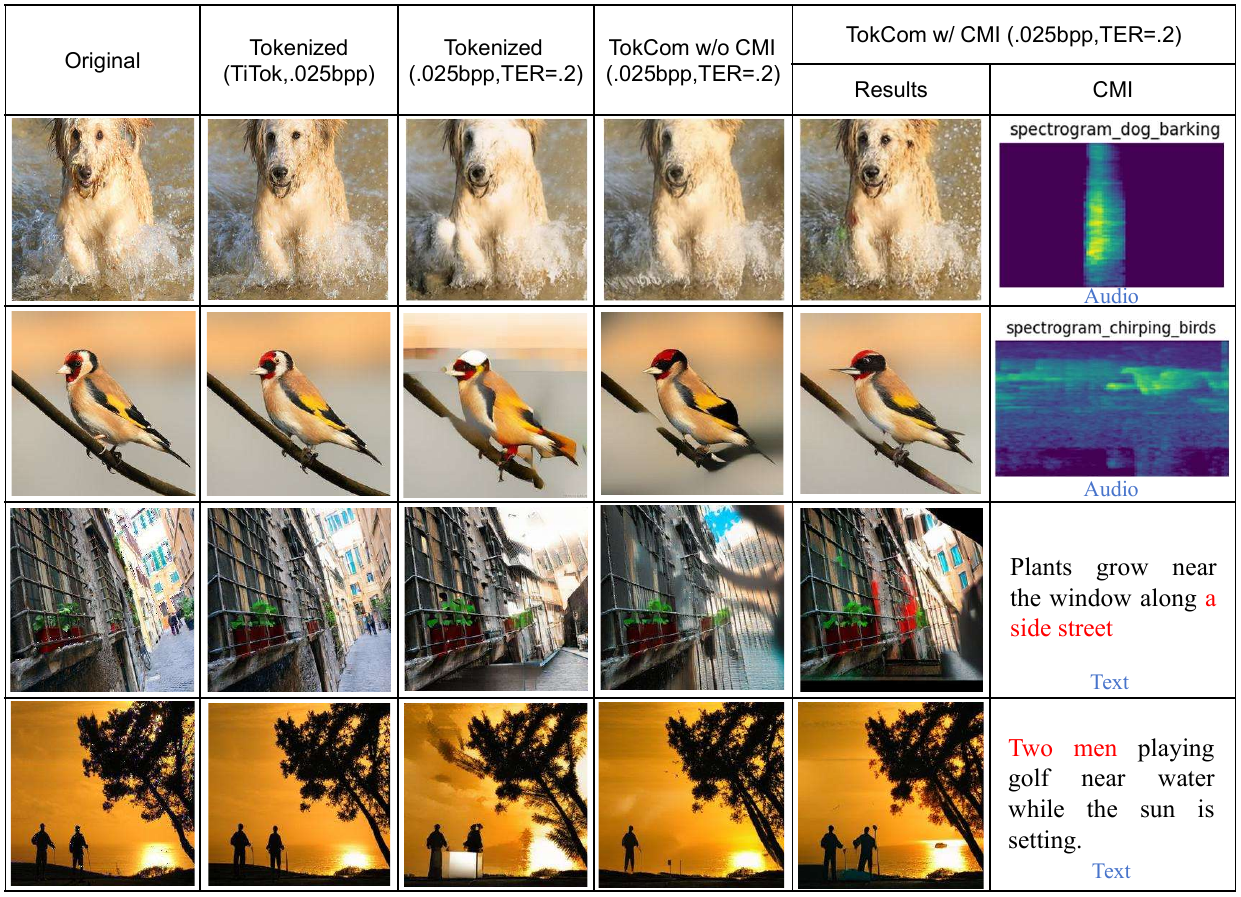}}
\captionsetup{font={footnotesize, color = {black}}, singlelinecheck = off, justification = justified,name={Fig.},labelsep=period}
\caption{{Illustration of TokCom performance with audio and text-based CMI.}}
\label{fig_SimResults5}
\vspace{-4mm}
\end{figure}

\section{Open Problems and Future Research Directions} \label{Open}
\subsection{Efficient Tokenizer Design for TokCom}
Different tokenization schemes impact token predictability, and thereby the semantic compression performance, by balancing the input length versus entropy of the token distribution. For example, BPE tokenization reduces the length by merging frequently co-occurring tokens into larger tokens, which allows models to process more input data within a limited context window. However, this increases the token codebook size which can make token predictions more challenging. The balance between token size and input length, and hence design of efficient tokenizers, is crucial for TokCom efficiency. Another challenge is designing efficient unified tokenizers across different modalities. The primary difficulty lies in aligning the diverse representations of text, images, and other modalities, while ensuring that they can effectively interact in a shared embedding space. As multimodal models continue to scale, developing tokenizers that achieve this balance will be the key to unlocking the full potential of TokCom for efficient multi-modal GenSC.

{\subsection{TokCom Computational Complexity and Collaborative Inference}

According to the scaling laws, performance of LMs reliably improve with their size and computational complexity \cite{kaplan2020scaling}. The existing GFM/MLLMs are typically computationally complex, thereby posing challenges in their deployment in the TokCom framework. To tackle this, future research could explore design of collaborative device-edge-cloud inference schemes, where lightweight models can run on-device, while more complex inference is offloaded to edge or cloud servers equipped with high-performance larger models. The offloading strategy should optimize the trade off between computational complexity, latency, and the resulting semantic quality. The corresponding latency-performance tradeoffs should be derived for TokCom by considering factors such as model size, computational resources, channel conditions, and the heterogeneity in zero/few-shot performance of pre-trained GFM/MLLMs deployed on device-edge-cloud \cite{Mengmeng}. Another solution could be reasoning through collaboration between large and small models leveraging speculative decoding techniques \cite{JP}, which could be further studied in future research.}

{\subsection{TokCom Privacy and Security}

The use of tokens as communication units and the reliance on large pre-trained GFM/LLMs introduces potential vulnerabilities. Tokenized representations carry semantic payloads, and hence are more vulnerable to inference attacks. An adversary intercepting a TokCom flow can easier manipulate the transmitted content through tokens. The strong context information in token sequences may lead to leakage of sensitive information, e.g., about the communication intent. Future work could explore design of new encryption and privacy-preservation techniques in the token embedding space to mitigate these challenges. Moreover, using pre-trained GFM/MLLMs could be susceptible to adversarial attacks. Differential privacy schemes should be explored for pre-training to prevent leakage of any sensitive information from the training data. Pre-trained models may inherit or amplify biases present in their training data, which could adversely affect TokCom systems. The use of machine unlearning techniques could be explored to remove influence of any biased training data. As TokCom evolves, ensuring privacy and security at both the architectural and operational levels will remain paramount to its effective deployment in future wireless networks.}

{\subsection{6G Applications and Network Architecture}

Finally, future research could explore emerging 6G applications such as the wireless metaverse and immersive communications, extended/mixed reality (XR/MR), the internet of senses, and holographic teleportation, where TokCom can enable efficient intent-aware communications leveraging cross-modal context information. To enable this, design of semantic-aware network architecture should be explored to allow distributed GFM/MLLMs deployment and support dynamic Knowledge Base (KB) management across the network layers. Scalable resource allocation schemes for TokCom on the semantic-aware radio access networks \cite{S-RAN} should be developed to adapt not only to the channel conditions and device capabilities, but also to the multimodal context information.}

\section{Conclusions}\label{SEC_Conclusion}
In this paper, we introduced TokCom, a new framework motivated by the recent success of MLLMs, for leveraging cross-modal context information in generative SemCom. We also presented the key setups for efficient TokCom at various layers in future wireless networks and demonstrated the corresponding benefits in a typical image SemCom setup. TokCom opens up new avenues to develop innovative context-aware multimodal generative semantic communication schemes, driving the evolution of future wireless networks.


\end{document}